\documentclass[twocolumn]{jpsj2} %% two-column layout
\newcommand{\Romannum}[1]{\uppercase\expandafter{\romannumeral#1}}

\title{Evidence for $s$-wave superconductivity with antiferromagnetic fluctuations in filled skutterudite LaFe$_{4}$P$_{12}$: $^{139}$La and $^{31}$P-NMR studies}
\author{
Yusuke. \textsc{Nakai}$^{1}$\thanks{E-mail address: nakai@scphys.kyoto-u.ac.jp},
Kenji. \textsc{Ishida}$^{1}$\thanks{E-mail address: kishida@scphys.kyoto-u.ac.jp},
Daisuke. \textsc{Kikuchi,}$^{3}$
Hitoshi. \textsc{Sugawara,}$^{2}$\\
and Hideyuki. \textsc{Sato,}$^{3}$}
\inst{$^{1}$Department of Physics, Graduate School of Science, Kyoto University, Kyoto 606-8502, \\
$^{2}$Faculty of Integrated Arts and Sciences, Tokushima University, Tokushima 770-8502,\\
$^{3}$Graduate School of Science, Tokyo Metropolitan University, Hachioji, 192-0397\\}

\abst{We have performed $^{139}$La,$^{31}$P-NMR studies on a filled skutterudite superconductor LaFe$_{4}$P$_{12}$ with a critical temperature $T_c=4.1$ K.
In the normal state, the presence of antiferromagnetic (AFM) fluctuations at finite wave vectors is suggested from the relation between the nuclear spin-lattice relaxation rate $1/T_1$ and the Knight shift. In the superconducting (SC) state, the distinct coherence peak was observed just below $T_c$, and is easily suppressed by the applied field.
An exponential decrease of $1/T_1$ was observed, suggestive of the isotropic superconducting gap with $2\Delta/k_{\rm B}T_c =3.8$. 
Besides, we observed the decrease of the Knight shift, indicative of the singlet-pair formation below $T_c$.
These results are clear evidence that LaFe$_{4}$P$_{12}$ is a rare $s$-wave superconductor with significant AFM fluctuations.  
}

\kword{skutterudite, superconductivity, Knight shift, NMR}

\begin{document}
\maketitle
%%%%%%%%%%%%%%%%%%%%%%%%%%

Filled skutterudite systems with the general formula $Ln$T$_4$X$_{12}$ ($Ln$ = lanthanide, T = Fe, Ru, Os, X =P, As, Sb) possess a wide variety of interesting physical properties. Among them, $Ln$-Fe$_4$P$_{12}$ have attracted much attention since PrFe$_4$P$_{12}$ exhibits heavy fermion behavior in magnetic fields adjacent to an antiferroquadrupole (AFQ) ordering phase\cite{H.Sato, AokiPRB, Sugawara}. The mechanism of the heavy fermion behavior has been now studied intensively from both theoretical and experimental points of view. 
Depending on $Ln$, $Ln$-Fe$_4$P$_{12}$ compounds shows various ground states, e.g.
LaFe$_4$P$_{12}$ is a superconductor, CeFe$_4$P$_{12}$ is a semiconductor, NdFe$_4$P$_{12}$ shows a ferromagnetic transition and so on.
In order to clarify the properties of $Ln$-Fe$_4$P$_{12}$ with 4f electrons, it is important to investigate the characteristic electronic structure of Fe$_4$P$_{12}$ cage in LaFe$_4$P$_{12}$ without $4f$ electrons.

LaFe$_4$P$_{12}$ is one of the rare superconductors containing iron with the superconducting (SC) transition temperature $T_c \approx 4.1$K. LaFe$_4$P$_{12}$ has a moderately high electronic specific heat coefficient of 57 mJ/K$^2$mol\cite{G.P.Meisner}, and its electrical resistivity shows a metallic temperature dependence.
M{\"o}ssbauer experiments\cite{Shenoy} and magnetic susceptibility measurements\cite{G.P.Meisner} on LaFe$_4$P$_{12}$ show that the iron does not have a local moment.
The Fermi surface seems as a distorted cube and the volume is almost a half of the BZ with a sharp peak in the 
density of states (DOS), indicating that the nesting with ${\bf Q} = (1, 0, 0)$ is likely in the filled skutterudite compounds\cite{SugawaraJPSJ}. 
In fact, this nesting vector affects physical properties in the filled skutterudite compounds. For instance, structural phase transitions with ${\bf Q} = (1, 0, 0)$ are observed below a metal-insulator transition temperature $T_{\rm MI}$ in PrRu$_4$P$_{12}$\cite{Lee}  and an AFQ ordering temperature $T_{\rm AFQ}$ in PrFe$_4$P$_{12}$\cite{Iwasa} respectively.  

In this paper, we report microscopic properties of LaFe$_4$P$_{12}$ in the normal and SC states obtained by $^{139}$La and $^{31}$P-nuclear magnetic resonance (NMR) measurements.
The results of NMR shift $K$ and spin-lattice relaxation rate $1/T_1$ are presented.
From these results, we conclude that LaFe$_4$P$_{12}$ is a rare $s$-wave superconductor with significant antiferromagnetic (AFM) correlations in the normal state.

NMR measurements were performed using powdered samples from single crystals which were grown by the tin-flux method described in detail elsewhere.\cite{Sugawara}. 
The NMR measurements were performed in the temperature range between 1.4 K and 250 K using a conventional pulsed NMR technique. The nuclear spin-lattice relaxation time of $^{139}$La and $^{31}$P ($^{139}T_1$ and $^{31}T_1$) was measured by the saturation-recovery method, and was uniquely determined by a single component in the whole measured temperature range.  

Since the La nucleus (nuclear spin $I = 7/2$) is situated at a cubic-symmetry site, we observe a single sharp NMR line without a quadrupolar effect, the linewidth of which is approximately 20 kHz. The $^{139}$La NMR shift $^{139}K$ is then precisely determined from the peak position.
The spin part of $^{139}K$ ($^{139}K_s$) was estimated from the plot of $^{139}K$ against the susceptibility $\chi$ (the $K$-$\chi$ plot). A linear relationship holds in the plot above 130 K but deviates at lower temperatures. This is due to the Curie behavior observed in $\chi$, which is ascribed to a tiny amount of Fe in the Sn flux and/or some impurity phase.  

Figure 1 shows the $T$-dependence of the spin part of $^{139}K$ ($^{139}K_s$), along with $^{139}(1/T_1T)$. 
$^{139}K_s$ is almost temperature-independent, while $^{139}(1/T_1T)$ increases with decreasing temperature, and is nearly constant in the temperature range below 40 K.

The $^{139}K_s$ is proportional to the spin susceptibility which is not affected by the impurity effect. From the $K$-$\chi$ plot, the hyperfine coupling constant at the La site is estimated to be $^{139}A_{\rm hf} = 9.40$ (kOe/$\mu_{\rm B}$). Thus the spin susceptibility at low temperatures is estimated to be $\chi_{s} = N_{\rm A}\mu_{\rm B}\frac{^{139}K_s}{^{139}A_{\rm hf}} = 7.11\times10^{-4}$ emu/mol. The DOS at the Fermi surface is also calculated to be $N(E_{\rm F}) = \chi_s / \mu_{\rm B}^2 = 22.0$ state/eV, which is about 1.7 times larger than that obtained from the band calculation\cite{SugawaraJPSJ}. From the comparison with the electronic specific heat coefficient $\gamma$ (57 mJ/K$^2$mol\cite{G.P.Meisner}), the Wilson ratio $R_{\rm W} = \frac{\chi_s/\chi_{\rm band}}{\gamma/\gamma_{\rm band}} = 0.91$ is estimated, which implies that the electron-phonon enhancement is larger than the spin enhancement.

In the presence of spin fluctuations, the ratio $R$ between an experimental value of $(1/T_1TK_s^2)$ and the Korringa value of non-interacting electron system $S_0$ ($R = S_0/T_1TK_s^2 $) provides an important information about magnetic correlations, where $S_0 = (\gamma_e/\gamma_n)^2(\hbar/4\pi k_{\rm B}) = (T_1TK_s^2)_0$.
The $\gamma_e$ and $\gamma_n$ are the electronic and nuclear gyromagnetic ratio, respectively. 
Since $1/T_1T$ is proportional to the summation of the $q$-dependent dynamical susceptibility over the $q$ space, it is enhanced by either ferromagnetic (FM) or antiferromagnetic (AFM) spin correlations, while only FM correlations strongly enhance the Knight shift. 
Therefore, the value of $R$ is larger (smaller) than unity when AFM (FM) correlations exist.
Using the experimental values of $^{139}(1/T_1T)$ and $^{139}K_{\rm s}$, $R$ is estimated as shown in the inset of Fig. 1. The $R$ is constant with $\sim 1.6$ down to 150 K and increases up to $\sim 3$ at lower temperatures, which indicates the development of AFM fluctuations with a finite $q$ vectors far from $q = 0$ in low temperatures.  

In a metallic compound with AFM fluctuations, the self-consistent renormalization (SCR) theory of spin fluctuations predicts $1/T_1T\propto \chi({\bf Q})^{1/2}$ for a 3-dimensional system, where $T$-dependence of the staggered susceptibility is given by $\chi({\bf Q})\propto 1 / (T+\theta)$. 
Actually, the $1/T_1T$ data above 40 K can be fit as $1/T_1T = a / \sqrt{T+\theta}$ with $a =$ 2.4 sec$^{-1}$K$^{-1/2}$ and $\theta = 35$ K (solid line in Fig. 1).
The relation of $1/T_1T\propto \chi({\bf Q})^{1/2}$ down to 40 K and the value $R\approx 2.5$ at 50 K suggests the development of the AFM correlations in LaFe$_4$P$_{12}$. 
As far as we know, the present NMR results are the first evidence that AFM fluctuations are present in LaFe$_4$P$_{12}$.
Below 40K, $1/T_1T$ deviates from the SCR relation and becomes constant (Korringa relation) below 30K.
 The deviation from the SCR theory below 40K is interpreted as a crossover to the 
Fermi liquid behavior. In fact, it is reported that the resistivity follows the $T^2$ relation below 25K\cite{H.Sato}, which is a characteristic feature of the Fermi liquid state.
%************************Fig.1***************************************
\begin{figure}[tb]
\begin{center}
\includegraphics[width=8cm]{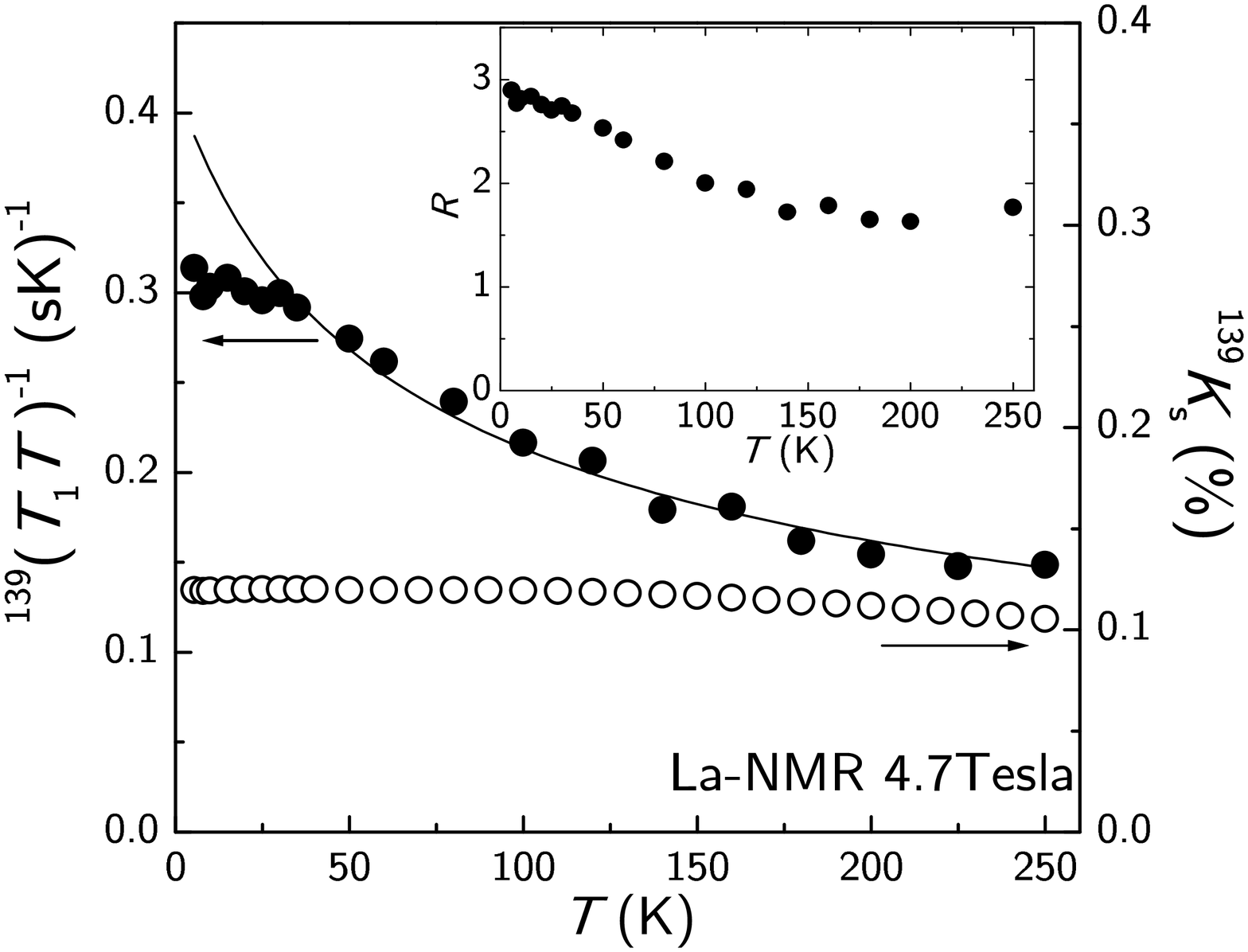}
\end{center}
\caption{The temperature dependence of $^{139}(T_1T)^{-1}$ and the Knight shift at the La site. 
The solid line is based on the SCR theory described in the text. Inset: the temperature dependence of $R$.}
\label{f1}
\end{figure}
%***********************************************************************
%************************Fig.2***************************************
\begin{figure}[tb]
\begin{center}
\includegraphics[width=7cm]{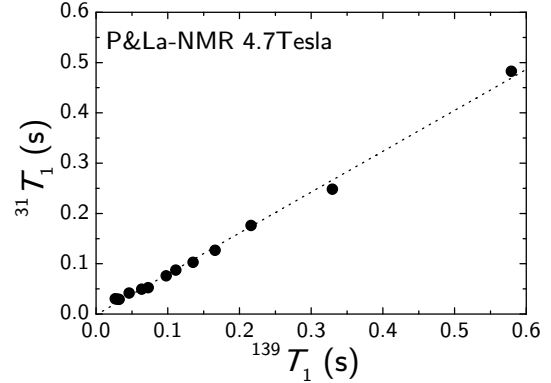}
\end{center}
\caption{Spin-lattice relaxation time at the P site vs that at the La site with temperature as an implicit parameter 
between 5.5K and 250K. }
\label{}
\end{figure}
%***********************************************************************

In the SC state, we measured $1/T_1$ mainly at the P site. This is because a signal to noise ratio of the P-NMR signal is better than that of the La signal due to the larger gyromagnetic ratio. To know the relation between $1/T_1$ at the P and La sites, $^{31}T_1$ in the normal state is plotted against $^{139}T_1$ as shown in Fig. 2. \cite{Ishida} 
A linear relation holds between $^{31}T_1$ and $^{139}T_1$, indicating that both $1/T_1$'s are dominated by the same dynamical susceptibility.   
Since $q$-dependence is negligible for a metallic compound without strong correlations, the hyperfine coupling constant at the $^{31}$P site is estimated to be $^{31}A_{\rm hf} = 3.64$ (kOe/$\mu_{\rm B}$) from the relation of $^{31}T_1/^{139}T_1 = (^{139}\gamma_n \cdot ^{139}A_{hf}/^{31}\gamma_n \cdot ^{31}A_{hf})^2 = 0.81 $ and the value of $^{139}A_{\rm hf}$.

%************************Fig.3***************************************
\begin{figure}[tb]
\begin{center}
\includegraphics[width=6cm]{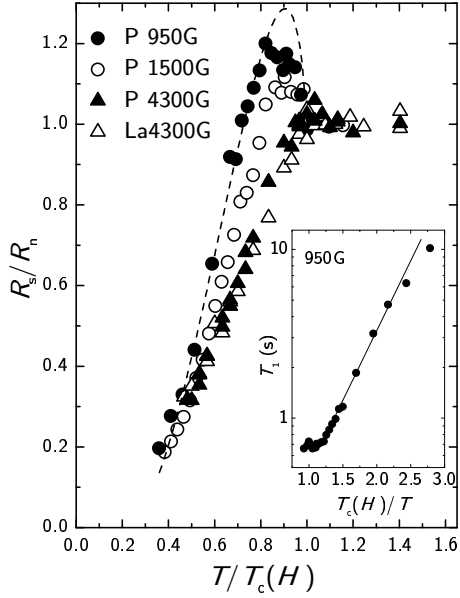}
\end{center}
\caption{$R_{\rm s}/R_{\rm n}$ vs $T/T_c(H)$ under various magnetic fields. The broken line shows the calculation on the basis of the 
$s$-wave model with $2\Delta(0)/k_{\rm B}T_c=3.8$. 
Inset : An Arrhenius plot of $T_1$ vs $T_c(H)/T$ obtained by $^{31}$P-NMR. The solid line represents the relationship
$T_1\propto \exp{(\frac{\Delta(0)}{k_{\rm B}T})}$ with $2\Delta(0)/k_{\rm B}T_c=3.8$.  }
\label{f2}
\end{figure}
%***********************************************************************
Let us discuss $^{31}(1/T_1)$ and $^{139}(1/T_1)$ results in the SC state. 
The properties of the SC gap are obtained from the temperature dependence of $1/T_1$ in the SC state.
The inset of Fig. 3 shows an Arrhenius plot of $T_1$ vs. $T_c(H)/T$, where $T_1$ was measured in a small field of 950 Oe. Obviously, experimental $T_1$ data obey the exponential behavior as $T_1\propto \exp{(\Delta(0)/k_{\rm B}T)}$ below $T_c$. The slope of the solid line indicates that $2\Delta(0)/k_{\rm B}T_c=3.8$. This is in good agreement with the value of the BCS theory in the weak-coupling regime, $2\Delta(0)/k_{\rm B}T_c=3.5$. The main panel of figure 3 shows the temperature dependence of $R_s/R_n [=(1/T_1)_s/(1/T_1)_n]$, where $(1/T_1)_s$ [$(1/T_1)_n$] corresponds to the data in the SC [normal] state. 
$^{139}(1/T_1)$ and $^{31}(1/T_1)$ measured in a magnetic field of 4300 Oe exhibits no clear coherence peak just below $T_c$, however, with lowering external magnetic fields, we observed a distinct coherence peak just below $T_c$. 
This result is a strong evidence for a conventional $s$-wave superconductivity with a finite SC gap.

If $H_{c2}(0) \sim 17$ kOe \cite{Sato} is taken into account, it was revealed that the coherence peak is easily suppressed by a much smaller magnetic field than $H_{c2}$. It is likely that the suppression of the coherence peak by an applied field occurs in YFe$_4$P$_{12}$, in which a small coherence peak is observed at an applied magnetic field of 3.1 kOe, but no peak in 11.7 kOe.\cite{Magishi} 
On the other hand, it should be noted that $1/T_1$ in LaOs$_4$Sb$_{12}$ by an Sb-nuclear quadrupole resonance (NQR), which was measured in a zero field, shows a distinct coherence peak just below $T_c$. \cite{Kotegawa} 
These experimental facts suggest that it is necessary to measure $1/T_1$ in the lowest magnetic fields as possible or in zero-magnetic field using NQR technique in order to detect the coherence peak which is a characteristic feature of an $s$-wave superconductor.

The experimental $R_s/R_n$ is reproduced by a calculation based on a theoretical model of a typical $s$-wave superconductor using the SC DOS $N_s(E)$. \cite{Hebel}
$R_s / R_n$ is expressed  as follows,
\[\frac{R_s}{R_n} \propto \frac{2}{k_BT_c}\int_0^{\infty}{(N_s(E)^2+M(E)^2)f(E)(1-f(E))}dE
\] 
where $M(E)$ and $f(E)$ are the so-called {\it anomalous} DOS arising from the coherence effect of the scattering inherent in a spin-singlet SC state and the Fermi-distribution function, respectively \cite{MacLaughlin}. 
The coherence peak originates from the presence of $M(E)$ and the singularity of $N_s(E)$ at $E = \Delta$. 
In the calculation, the phenomenological energy broadening function in $N_s(E)$ and $M(E)$ is assumed to be of rectangle type of the function with a width $2\delta$ and height $1/2\delta$ to suppress the singularity at $E = \Delta$, and BCS-type temperature dependence on $\Delta(T)$ is employed. 
The broken line  below $T_c$ in Fig. 3 shows the results of a calculation using $2\Delta(0)/k_{\rm B}T_c=3.8$ and $\delta/\Delta(T) = 0.25$. Although the calculation cannot reproduce the behavior just below $T_c$, we obtain good agreement with the experimental results below 0.8 $T_c$. 
%************************Fig.4***************************************
\begin{figure}[tb]
\begin{center}
\includegraphics[width=5cm]{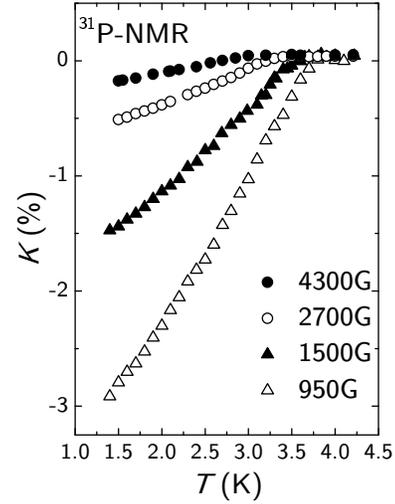}
\end{center}
\caption{The temperature dependence of $^{31}K$ under various magnetic fields.}
\label{f3}
\end{figure}
%***********************************************************************

Next, we show experimental results of NMR shift measured in the SC state.
Figure 4 shows the temperature dependence of $^{31}K$ in various fields. 
We found that a decrease of $^{31}K$ becomes larger when applied magnetic fields is lowered, which indicates that the SC diamagnetic shift $K_{\rm dia}$ is dominant below $T_c$. This is reasonable because the spin part of the Knight shift $^{31}K_s$ estimated from $K$-$\chi$ plot is approximately 0.1\% which is much smaller than $K_{\rm dia}$. 
When the vortices form a triangular lattice, $K_{\rm dia}$ is given by\cite{P.G} 
\begin{equation}
\label{dia}
K_{\rm dia}(T)=-\frac{1-N}{H}\frac{\Phi_0}{4\pi\lambda(T)^2}\ln \left(\frac{0.381 e^{-\frac{1}{2}}}{\xi(T)}\sqrt{\frac{2\Phi_{0}}{\sqrt3H}}\right) 
\end{equation}
where $\Phi_0$ is the flux quantum, $\lambda$ the magnetic penetration depth, $\xi$ the coherence length, and $N(=1/3)$ the 
demagnetization factor.
%************************Fig.5***************************************
\begin{figure}[tb]
\begin{center}
\includegraphics[width=8cm]{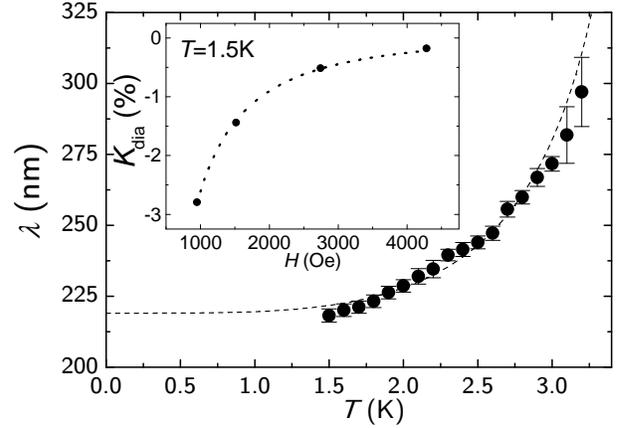}
\end{center}
\caption{The penetration depth obtained from the fitting of the inset data. The dashed line represents $\lambda(T)=\lambda(0)/\sqrt{1-(T/T_c)^4}$. Inset: the magnetic field dependence of $K\sim K_{\rm dia}$, in which 
the dotted line represents the theoretical curve written by Eq.(1). }
\end{figure}
%***********************************************************************
If we assume that $^{31}K$ is ascribed to SC diamagnetic shift, {\it i.e.}, $K \approx K_{\rm dia}$, 
$\lambda(1.5{\rm K})$ is estimated to be 220\AA ~ from the fitting the Eq.(1) to the experimental data of $K_{\rm dia}$ as shown in the inset of Fig. 5. Here the experimental value of $\xi(0)\approx 10$ nm \cite{Sato} and its temperature dependence of $\xi(T)=\xi(0)\sqrt{\frac{1+(T/T_c)^2}{1-(T/T_c)^2}}$ are employed. 
This fitting at each temperature  yields the temperature dependence of $\lambda(T)$ which is plotted in the main panel of Fig. 5.
Using $\lambda(T)=\lambda(0)/\sqrt{1-(T/T_c)^4}$, we estimate the penetration depth $\lambda(0) \approx 220$ nm.
This value is in reasonable agreement with the previously reported value $\lambda(0)\sim$ 140 nm \cite{Sato}, which supports the dominance of $K_{\rm dia}$ in the observed $K$.

Considering that the spin part of the Knight shift $^{31}K_s$ obtained by $K$-$\chi$ plot is approximately 0.1\%, it is difficult to observe the decrease of $K_{\rm s}$ due to the large diamagnetic shift in the SC state.
However, we succeeded in detecting the decrease of the $K_{\rm s}$ from the measurement of two NMR active nucleus in this compound, as shown below.
%Kp-KLa% 
The total NMR shift $^{\alpha}K$ for nucleus $\alpha$ ($\alpha =$139(La), 31(P)) is given by
\[^{\alpha}K(T, H) = ^{\alpha}K_{\rm s}(T) + ^{\alpha}K_{\rm orb} + K_{\rm dia}(T, H) .\]
$^{\alpha}K_{\rm orb}$ is the orbital shift which is temperature independent. $K_{\rm dia}$ is ascribed to the screening currents which is the same for all nuclear species in the sample because it is a bulk phenomenon.
Thus, the difference between $^{31}K(T)$ and $^{139}K(T)$ does not include the $K_{\rm dia}(T, H)$ and expressed as;
\begin{eqnarray*}
^{31}K(T) - ^{139}K(T) &=& (^{31}K_{\rm s}(T) - ^{139}K_{\rm s}(T)) + {\rm const.} \\
                      &=& \frac{(^{31}A - ^{139}A)}{N_{\rm A}\mu_{\rm B}}\chi_{\rm s}(T) + {\rm const.} 
\end{eqnarray*}
Here, the relation of $^{\alpha}K_{\rm s}(T) = (^{\alpha}A/N_{\rm A}\mu_{\rm B})\chi_{\rm s}(T)$ is used.

%************************Fig.6***************************************
\begin{figure}[tb]
\begin{center}
\includegraphics[width=8cm]{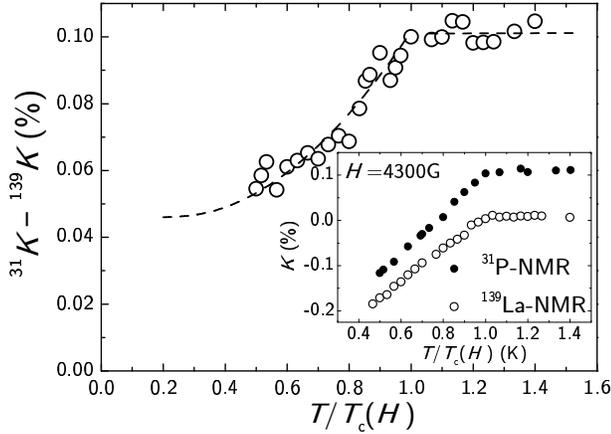}
\end{center}
\caption{$^{31}K-^{139}K$, vs $T/T_c(H)$. Dotted line represents the Yosida function with the same 
gap energy with those used in the analysis of $1/T_1$ below $T_c$. Inset: $^{31}K$ and $^{139}K$ vs $T/T_c(H)$ under 4300Gauss.} 
\label{f5}
\end{figure}
%***********************************************************************
Figure 6 shows the temperature dependence of $^{31}K - ^{139}K$ which is proportional to $\chi_{\rm s}(T)$ in the SC state. 
The temperature dependence of $\chi_{\rm s}$ in a spin-singlet SC state is described by the Yosida function.\cite{Yosida}
The broken curve in Fig.~6 is the calculation based on the Yosida function with the same gap value obtained by the $1/T_1$ analysis. The fairly good agreement is obtained. The temperature dependence of $1/T_1$ and Knight shift in the SC state are well understood by the BCS SC model with $2\Delta(0)/k_{\rm B}T_c=3.8$, which strongly evidences
 that LaFe$_4$P$_{12}$ is a spin-singlet $s$-wave superconductor. 

Finally, we comment on the magnetic fluctuations in LaFe$_4$P$_{12}$. The magnetic properties on the filled skutterudite compounds by non-magnetic ions such as 
La and alkalies have drawn a great deal of attention recently since the itinerant ferromagnetic order and nearly ferromagnetic state was discovered in $A$Fe$_4$Sb$_{12}$ ($A$ $=$ Na and K)\cite{Jasper} and $M$Fe$_4$Sb$_{12}$ ($M$ $=$ Ca, Sr, Ba)\cite{Matsuoka, Matsumura}, respectively and ferromagnetic correlations are suggested in LaFe$_4$Sb$_{12}$\cite{Magishi2}. In LaFe$_4$Sb$_{12}$, the band calculation pointed out that the sharp peak of the density of states formed by the Fe-$3d$ electrons is close to the Fermi level, which is the origin of the magnetism. 
Contrary to this, the sharp peak from the Fe-$3d$ electrons in LaFe$_4$P$_{12}$ is below the Fermi level and 
the magnetism from the Fe-$3d$ electrons seems to be weaker than that in LaFe$_4$Sb$_{12}$. 
Rather the observed AFM fluctuations seem to originate from the Fermi surface properties as discussed below. 
According to the dHvA experiments, it is reported that Fermi surfaces consist of multi-bands; the multiply-connected Fermi surface with $m^*=22m_0$ denoted as 48-th band and the nearly spherical Fermi surface with $m^*=2.2m_0$ denoted as 47-th band in the ref.6. It is considered that the AFM fluctuations originates from the nesting properties of the 48-th Fermi surface with ${\bf Q} = (1,0,0)$, which is suggested by the band calculation\cite{SugawaraJPSJ}. Besides, Harima and Takegahara reported that the nesting properties of the 48-th Fermi surface is suppressed in LaFe$_4$Sb$_{12}$ due to the rising of Fe-$3d$ band\cite{Harima}. Therefore, it seems that 
the position of Fe-$3d$ level is important for determining the magnetic properties in LaFe$_4$X$_{12}$ (X $=$ P, Sb).

In summary, we have reported that $^{31}$P and $^{139}$La-NMR measurements on a filled skutterudite superconductor LaFe$_4$P$_{12}$.
In the normal state, $^{139}(1/T_1T)$ and $^{139}K$ suggest that AFM spin fluctuations with $q$-wave vectors far from $q = 0$ develops down to 40 K.
In the SC state, $1/T_1$ shows an exponential decrease below $T_c$ with the isotropic energy gap $2\Delta/k_{\rm B}T_c =3.8$.

Furthermore, we succeeded in detecting the decrease of the spin susceptibility, indicative of a spin singlet pair.
From these results, we conclude that LaFe$_4$P$_{12}$ is an $s$-wave superconductor with the notable AFM fluctuations, which originate from the nesting properties of the Fermi surface.

We thank D. E. MacLaughlin, H. Harima, Y. Maeno, K. Kitagawa, H. Murakawa and Y. Ihara for valuable discussions and experimental support.
This work was supported by the Grant-in-Aid for the 21st Century COE "Center for Diversity and Universality in Physics" from the Ministry of Education, Culture, Sports, Science and Technology (MEXT) of Japan, and by the Grants-in-Aid for Scientific Research from Japan Society for Promotion of Science (JSPS),
and by a Grant-in-Aid for Scientific Research in Priority Area "Skutterudite" (No.16037208, 15072206).

\end{document}